\documentclass[]{spie}  %>>> use for US letter paper
\usepackage[dvips]{graphicx}

\newcommand{\apj}{\em Astrophys. J.}
\newcommand{\mnras}{\em M.N.R.A.S.}
\newcommand{\apjl}{\em Astrophys. J. Lett.}

\newcommand{\pasp}{\em P.A.S.P.}
\newcommand{\nat}{\em Nature}
\newcommand{\prd}{\em Phys. Rev. D.}

\def\coord2#1#2#3#4{{{#1}$^h$ {#2}$^m$, {#3}$^\circ$ {#4}$^\prime$}}

\def\kms{{km~s$^{-1}$}}

\def\simgt{\ {\raise-.5ex\hbox{$\buildrel>\over\sim$}}\ } % < OR APPROX EQUAL TO
\def\simlt{\ {\raise-.5ex\hbox{$\buildrel<\over\sim$}}\ } % > OR APPROX EQUAL TO

\def\pp{\noindent\parshape 2 0truecm 16.25truecm 1truecm 15truecm}
\def\apjref#1;#2;#3;#4 {\par\pp#1, {\it #2}, {\bf #3}, #4}
\def\jref#1;#2 {\par\pp#1}

\newcommand{\etal}{et al. }
\title{Science Objectives and Early Results of the  DEEP2 Redshift Survey }

\author{Marc Davis\supit{a}, 
Sandra M. Faber\supit{b}, 
Jeffrey A. Newman\supit{a},  
Andrew C. Phillips\supit{b}, \\
R.S. Ellis\supit{c},
C.C. Steidel\supit{c},
C. Conselice\supit{c}, \\
A. L. Coil\supit{a},
D. P. Finkbeiner\supit{d},
D. C. Koo\supit{b},
P. Guhathakurta\supit{b}, 
B. Weiner\supit{b}, R. Schiavon\supit{b}, C Willmer\supit{b},\\
N. Kaiser\supit{e}, G. Luppino\supit{e}, G. Wirth\supit{f},
A. Connolly\supit{g}, P. Eisenhardt\supit{h}\\
M. Cooper\supit{a}, B. Gerke\supit{a}
\skiplinehalf
\supit{a} University of California, Berkeley, CA,\\
\supit{b} Lick Observatory, University of California, Santa Cruz, CA,\\
\supit{c} Caltech, Pasadena, CA, \\
\supit{d} Princeton University, Princeton, NJ, \\
\supit{e} University of Hawaii, Hononlulu, HI \\
\supit{f} Keck Observatory, Waimea, HI \\
\supit{g} Carnegie Mellon University, Pittsburgh, PA \\
\supit{h} Jet Propulsion Laboratory, Pasadena, CA }

\authorinfo{(Send correspondence to M.D. E-mail
marc@astron.berkeley.edu)}

\begin{document} 
\maketitle              % typesets the title of the contribution

%%%%%%%%%%%%%%%%%%%%%%%%%%%%%%%%%%%%%%%%%%%%%%%%%%%%%%%%%%%%% 

\begin{abstract} The DEIMOS spectrograph has now been installed on the
Keck-II telescope and commissioning is nearly complete.  The
DEEP2 Redshift Survey, which will take
approximately 120 nights at the Keck Observatory over a
three year period and has been designed to utilize the power of 
DEIMOS, began in the summer of 2002.  The multiplexing power and high
efficiency of DEIMOS enables us to target 1000 faint galaxies per
clear night.  
Our goal is to gather high-quality spectra of $\approx
60,000$ galaxies with $z>0.75$ in order to study the properties
 and large scale clustering of galaxies at $z \approx 1$.  The survey will
be executed at high spectral resolution, $R=\lambda/\Delta \lambda \approx
5000$, allowing us to work between the bright OH sky emission lines and to
infer linewidths for many of the target galaxies (for several thousand
objects, we will obtain rotation curves as well).  The linewidth data will
facilitate the execution of the classical redshift-volume cosmological
test, which can provide a precision measurement of the equation of state
of the Universe. 
%We will also obtain clean rotation curves for several
%thousand galaxies at $z>0.75$.  
This talk reviews the project, summarizes our science
goals and presents some early DEIMOS data.  
\end{abstract}

\keywords{Redshift Surveys, Large Scale Structure, Galaxy Properties, 
         Cosmology}

\section{INTRODUCTION}

Our theoretical understanding of large scale structure and galaxy
formation is well advanced, yet many crucial questions remain.
Studies of structure depend on objects that can be seen, namely
galaxies, whereas the raw medium from which galaxies formed is a
mixture of both dark and baryonic matter.  It has become clear that
the formation of visible galaxies in the universe is highly uneven and
that galaxies are a ``biased'' tracer of the underlying mass.  Thus,
to study structure in the universe at high redshift, we must be able
to predict how the universe converted matter into luminous objects,
requiring a thorough understanding of the physics of galaxy formation
and evolution; but those processes depend on the underlying
cosmological parameters and structure which we would also like to study!

Untangling galaxy evolution from cosmological evolution is extremely
difficult with studies restricted to the local universe, which provide us
with only a snapshot of the end result of galaxy formation, or those
which include only a small number of very distant objects. Our ability to
draw conclusions from galaxies in the Hubble Deep Field, for instance,
is limited not only by the relatively small number of objects in the
field that are bright enough to be studied from the ground, but also
by the intrinsic spatial correlations between the galaxies, which
causes fluctuations of measurements performed over a small volume to be much
greater than simple Poisson statistics would suggest.  However, the
combination of a statistically robust, large-volume, high-redshift
sample with studies of the present-day universe should allow us to
untangle the properties of galaxies from studies of large-scale
structure, and simultaneously provide great amounts of information on
galaxy formation and evolution.

The DEEP2 (DEEP Extragalactic Evolutionary Probe 2) Redshift Survey
has been designed to produce such a dataset: one sufficiently rich
both to refine our knowledge of fundamental cosmology and to challenge
future galaxy formation models.  The DEEP2 collaboration plans to
obtain spectra of $\sim 60,000$ galaxies at high redshift using
DEIMOS, a new multi-object spectrograph recently commisioned on the
Keck Telescope. Details of DEIMOS are presented in this conference
 by Faber \etal \cite{faber02}.
 DEEP2 will provide a sample comparable in quality and
volume to local surveys such as the LCRS, and thereby constrain the
evolution of the properties of galaxies and of large-scale structure.
The full program is expected to occupy 120 nights of Keck time, spread
over a three year period.  By the end of 2002, 17 science nights will
have been allocated for this program. 
% (80 nights over 3+ years, beginning July 2002, are on schedule
%to be 
%be allocated by the UC TAC).   

\section{The DEEP2 Observing Plan}

%Now that DEIMOS is operational, we have commenced the  DEEP2 project, 
%a survey of faint
%galaxies designed to characterize galaxies and the galaxy distribution at
%a redshift $Z=1$. 
%The DEEP2 goal is to generate a sample of uniform
%quality data with a well defined selection criterion that will be suitable
%for many different analyses.  
This large effort has brought together Keck observers from UC, 
Caltech, and the University of Hawaii, in addition to outside
collaborators.  Team members with Keck access are M. Davis, S. Faber, D.
Koo, R. Guhathakurta, C. Steidel, R. Ellis, G. Luppino, and N. Kaiser.
Many nights of time on the Canada France Hawaii Telescope (CFHT) have
been dedicated to the wide-field imaging required for DEEP2, and 120
more at Keck will be required for the spectroscopic portion of the
survey. 

%Fields and details of the observing program have been provided in
%previous reviews (1).

\subsection{Fields and Photometry}
\begin{table}
\caption{1HS Fields Selected for the DEEP2 Redshift Survey}
\begin{center}
\begin{tabular}{|c|c|c|c|} \hline
RA & dec & (epoch 2000) & mask pattern \\ \hline
14$^h$ 17 & +52$^\circ$ 30  & Extended Groth Survey Strip  & 120x1 \\ \hline
16$^h$ 52 & +34$^\circ$ 55   & last zone of low extinction & 60x2\\ \hline
23$^h$ 30 & +0$^\circ$ 00   & on deep SDSS strip & 60x2 \\ \hline
02$^h$ 30 & +0$^\circ$ 00   & on deep SDSS strip & 60x2 \\ \hline
\end{tabular}
\end{center}
\end{table}

In order to overcome ``cosmic variance'', the excess fluctuations in
the distribution of galaxies (or galaxy clusters) due to their spatial
correlations, the DEEP2 Redshift Survey will target $\sim3.5$ deg$^2$ total
within four fields on the sky (listed in Table~1). 
 The fields were chosen as low
extinction zones that are continuously observable at favorable zenith 
angle from Hawaii over a six month interval. 
One field includes the extended Groth Survey strip (GSS), 
which has existing HST imaging and which will be the target
of very deep IR observations by SIRTF, and two of the fields are on the
equatorial strip which will be most deeply surveyed by the Sloan Digital Sky 
Survey (SDSS) project \cite{york}.  Each of these fields has been observed
by Luppino and Kaiser with the CFH12K camera in the B, R, and I bands.
For three of the fields, the imaging was obtained over three contiguous
pointings covering a  strip
of length $120'$ by $30'$ and oriented E-W;
however, the GSS field is oriented
along a line of constant ecliptic longitude, and therefore required 4
CFH12K pointings to cover it.

From the large photometric database resulting from these data, we have
used a simple cut in  $BRI$ color space
to select galaxies that should have redshift $z>0.75$.
Objects meeting this color cut and having magnitude $R_{AB}<24.1$ (for
the 1HS portion of the survey; q.v. below) are
then targeted for DEIMOS spectroscopy. 
% The deepest, most uniform 
%photometry is in the R
%band, and we have therefore chosen to select our spectroscopic 
%sample from the
%flux limited R band catalog.  DEIMOS is being used 
%to undertake a spectroscopic survey of those galaxies with $m_R(AB) \le
%24.1$ which  meet our color criteria.  
This photometric redshift preselection is a vital part of DEEP2.  Based on analysis of objects with known
redshifts within the DEEP2 photometric catalog, we find that $\sim 50\%$
of galaxies to the DEEP2 magnitude limit are at
redshift $z<0.75$; however, with a simple cut in color space we
can reduce this foreground fraction below 10\%, while eliminating only 
$\sim 3\%$ of high-redshift objects (mostly due to photometric errors
that cause them to cross the cut line).  
%{\bf Marc: do we want to include the figure? OK, we have space!} 
Applying such a photometric
preselection allows us to efficiently target galaxies at high
redshift, correspondingly reducing the number of spectra required to
perform a survey of distant regions of the universe.  With two masks
observed in early
commissioning, one designed with and one without a color cut, we have
already begun to test the efficacy of this method,
with the encouraging results shown in figure 1.  
%One mask was observed without a
%photo-z cut and another was observed with a cut; the resulting
%redshift distributions are consistent with expectations. 

\begin{figure}[t]
\begin{center}
\includegraphics[width=6in,height=4in,angle=90]{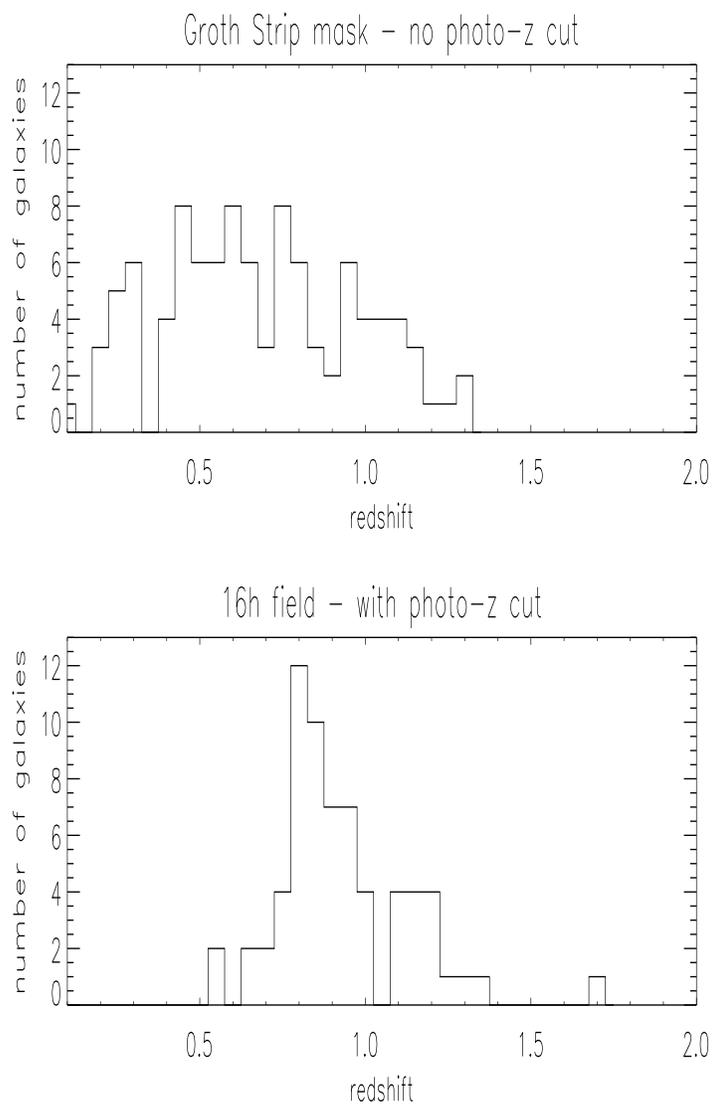}
\end{center}
\caption{The redshift distribution measured from data obtained during
June commissioning for 2 DEIMOS masks; one in the GSS without photo-z
preselection, and the other from the 16h field with photo-z preselection
designed to keep galaxies with $z>0.75$ .  Each mask was observed for
one hour.  The redshift
distribution observed in each case is fully consistent with
expectations; the fluctuations are consistent with the clustering
strength  expected for the survey. } 
\end{figure}

%At this relatively bright flux limit, 50\% of
%the galaxies have $Z<0.75$ \cite{lilly,cohen}; the photometric
%redshift preselection eliminates this foreground subsample, 
%allowing the DEEP2 project to focus its effort on the high-redshift
%Universe.  In the GSS region, the photo-z cut will not be made, and
%all galaxies in a region of 120' by 16' and $m_R < 24.0$ will be targeted
%for spectroscopy.  

Custom milled slitmasks must be designed and manufactured 
for each pointing of DEIMOS;
the masks for the DEEP2 Redshift Survey typically contain 120 slitlets for
galaxies, plus 4 boxes for alignment stars and 8 narrower slitlets in
blank areas used for non-local sky-subtraction.  Each mask covers a field
of view of 16' by 4', and the optical system reimages this mask onto
the CCD detector array of 8 MIT-LL CCD's, forming an 8k by 8k mosaic.

%\subsection{Choice of Grating and Spectral Resolution}
%A choice of gratings is available on DEIMOS, and 
%We have decided to use  that the
%workhorse grating for the DEEP2 survey will be the 900 lines/mm grating, 
%which has an anamorphic factor of 1.4.  This grating 
%will provide spectral coverage of 3500 $\AA$ in one setting.  If we use
%slits of width 0.75", they will project to a size of 4.6 pixels, or
%a wavelength interval of 2 $\AA$.  Thus the resolving power of the observations
%will be quite high, $R \equiv \lambda / \Delta\lambda =3700$.

The MIT-LL CCDs have  low fringing, high QE, and 
exceptionally low readout noise, 2.3 $e^-$; with 
the 1200 l/mm grating and a 1 arcsec slitwidth, the integration time
to become sky-noise limited in spectral 
regions between the OH night sky lines 
is only 10 minutes.  Figure 2 demonstrates the virtues of
working at high spectral resolution.
 The large number
of pixels in the dispersion direction allows high resolution with
substantial spectral range,  so that we can work between  the bright OH
sky emission lines while remaining sky-noise limited over the full
spectrum.  This 'OH
suppression' capability of DEIMOS suggests that improved SNR for faint
galaxies can be obtained by observing at high dispersion, and
later smoothing the data to lower resolution with proper accounting
of the variable sky noise.

\begin{figure}[t]
\begin{center}
\includegraphics[width=6in,height=6in]{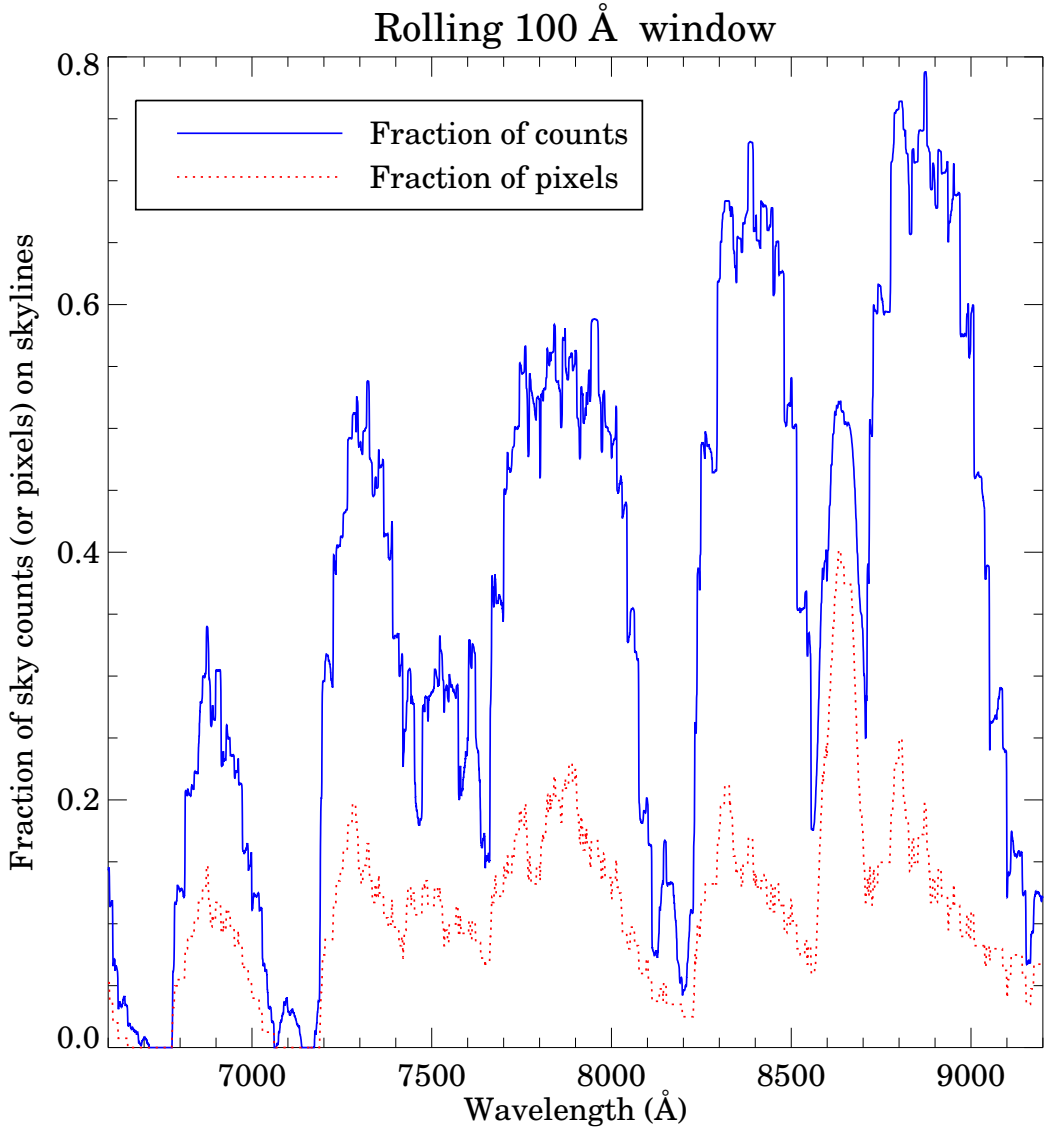}
\end{center}
\caption{The fraction of sky emission contributed by OH lines with
flux greater than twice the background level.  The solid curve shows
the fraction of the total flux, while the dotted  curve shows the
fraction of pixels contributing to this flux at resolution 5000.
Note that only 15-20\% of the pixels are contaminated by bright OH
emission; observing at high spectral resolution allows the use of 'OH
suppression' to greatly improve the typical signal-to-noise in our spectra. } 
\end{figure}

An added benefit of working at such high dispersion is that
corrections
for fringing are more stable.  Among other factors, the Fabry-Perot
fringing  that is
ubiquitous in CCD's in the near-IR has a fringe wavelength (in pixels)
which is proportional to the spectral dispersion.  If one seeks stable
flat fielding, it is important to hold the instrument's pixel to wavelength
registration to a small fraction of this wavelength.  In the case of 
DEIMOS, this period is $\sim 25$ Angstroms ($\sim 75$ pixels)
 with the 1200 l/mm grating,
and flexure compensation feedback limits the motion to $.1-.4$ pixels
for 180 degree rotation of the instrument.  Removing the effects of
fringing has been very straightforward in DEIMOS data.  If we were working at
resolution  500 rather than 5000, the requirements for instrument 
rigidity would be 10 times larger for equivalent removal of
fringing effects.

We are observing with one arcsec slitwidths and a
 1200 l/mm grating tilted so that the wavelengths
 6500-9100 $\AA$ are centered on
the detector, thus assuring that the 3727 \AA\ [OII] doublet is in range
for galaxies with $0.74 < Z < 1.44$.  At the planned spectral resolution, 
the velocity resolution will be 56 km/s.  
 Such high resolution is unconventional for faint objects but
has several benefits: (1) as shown above, 
most of the spectrum will be free of
terrestrial OH night-sky interference; (2) corrections for fringing
 are more stable, allowing us to achieve Poisson limited sky
 subtraction (see figure 3 below); (3) the O [II] doublet will be
resolved and will therefore yield a reliable redshift even if it is
the only feature observed; and (4) rotation velocities can be measured
down to $\sim 25$ \kms.  This last feature is unique to DEEP2 and sets
it apart from all other distant-galaxy surveys (indeed, most large
local surveys such as 2dF and SDSS have much lower velocity resolution
and no capability to measure rotation curves).  Work to date at UCSC
underscores the importance of this; [OII] linewidths measured from
high-resolution spectra do appear to be commensurate with
other indicators of potential well depth \cite{kobulnicky}.  The
signal-to-noise ratio in DEEP2 spectra should be adequate to measure
linewidths or even spatially-resolved rotation curves in roughly half
of our selected galaxies.

The original plan for the DEEP2 Redshift Survey was to use a lower dispersion
grating, 830 l/mm, for which DEIMOS would yield 3800 \AA\ of spectral
coverage at a resolution still sufficient to resolve the O [II] doublet.
However, for observations in the near IR, this grating must be tilted
nearly face-on to the DEIMOS camera.  Any light that
reflects from the CCD detector will pass through the camera and
grating a 2nd time; a fraction of this light will emerge in zeroth
order to re-enter the camera a third time, whereupon it will be
refocused at a mirror point on the detector.  These 'ghost images' are
in-focus and 
have an intensity of $\sim 0.5$\% of the original.  The
plethora of bright OH sky lines leads to serious ghost contamination
on the opposite side of the camera.  We have chosen to avoid dealing
with this issue by using the higher dispersion grating, which
has no ghosts in the near-IR.

\subsection{Observing Strategy--Target Selection }

In each of the four selected fields of Table 1, we will densely target a
region of 120' by 16' or 120' by 30' for DEIMOS spectroscopy. 
We intend to observe 120 separate slitmasks per field; each mask 
contains $\sim 120$ slitlets over 
%Each pointing of DEIMOS shall
%use a unique mask with slitlets cut over 
a region of size 16' by 4', with the slitlets mostly aligned along the
long axis of the mask, but with some tilted as much as $30^\circ$ to
track extended galaxies.
%Our goal is to select an average of 
% 120 slitlets per mask, selected from the
%those galaxies with $m_R(AB) \le 24.1$ and $z_{photo} > 0.75$.  
  The mean surface
density of candidate galaxies  exceeds the number of objects we
can select  by approximately 30\% (spectra of selected targets cannot be allowed to overlap on the CCDs).  
However, this will not cause problems with subsequent analyses if we take 
account of the positions of those galaxies for which we did not obtain 
spectroscopy.

In the Groth strip region, because of the interests of other 
collaborative scientific
projects, our plan is to construct 120 distinct masks each
offset from their neighbor 
by 1' and to select targets without regard to color, thereby not imposing
a photometric redshift preselection (this will allow us to sample
objects at $z<0.75$ and have the added benefit of providing an
internal test of the effects of our color selection).  Thus, any spot
on the sky will be found within 4 masks, giving every galaxy
 4 chances to appear
on a mask without conflict. In the other three survey fields, we plan 
to use the color selection, thus halving the source density of targets,
and
to step 2' between masks, giving a galaxy two chances to be selected without
conflict.  In these fields the masks will form a pattern of 60 by 2, covering
a field of 120' by 30'.  At $z \approx 1$, this field subtends a comoving
interval of $80~x~20 h^{-2}$ Mpc$^2$, and our redshift range translates to
a comoving interval of $\approx 800h^{-1}$ Mpc (assuming a flat Universe with 
$\Omega_\Lambda =0.7$).  Figure 1 shows that our photo-z selection
method is effective.

Note that this survey differs from the planned VLT/VIRMOS survey 
\cite{lefevre} in that a DEIMOS spectrum will occupy a full row of the CCD
array (8k pixels), so the VLT/VIRMOS  multiplexing will be higher than ours. 
The DEEP2 spectra will have resolution 20 times higher than planned for the
bulk of the VLT/VIRMOS project,  but our volume surveyed will be
smaller and thus more sensitive to cosmic variance.
%(partially because we are excluding objects with $Z<0.7$). 

Tests of slitmasks designed for DEEP2 began on the first night of
DEIMOS commissioning in early June, 2002.  On only the second night
of commissioning, precision mask alignment was achieved (this is a
difficult task, as tolerances are $<0.2$ arcsec over 15 arcmin) and
spectra were obtained of DEEP2 target galaxies.  An automated pipeline
now reduces the massive amounts of data obtained in DEEP2 observations
(140 MB per DEIMOS frame; each mask is observed multiple times to
eliminate cosmic rays, not to mention the calibration data required)
to sky-subtracted, coadded spectra.    We have a working pipeline for
reduction of the DEIMOS data, we and have no difficulty
producing fully reduced spectra within a day of observation.

An example of DEIMOS data is given in Figure 4, where we show
 a small piece of raw data for 
3 co-added 20 minute exposures taken through a DEEP2 science mask.  The
tilted sky lines and cosmic rays are prominent.  This same section of
the mask after pipeline processing is shown in figure 5.  For this 
figure each row has been aligned in wavelength by shifts 
in the spectral direction.  Note that the sky lines and cosmic rays
 have been cleanly removed, and that the underlying emission lines of the
target galaxies are prominent.

\begin{figure}[t]
\begin{center}
\includegraphics[width=6in,height=4in]{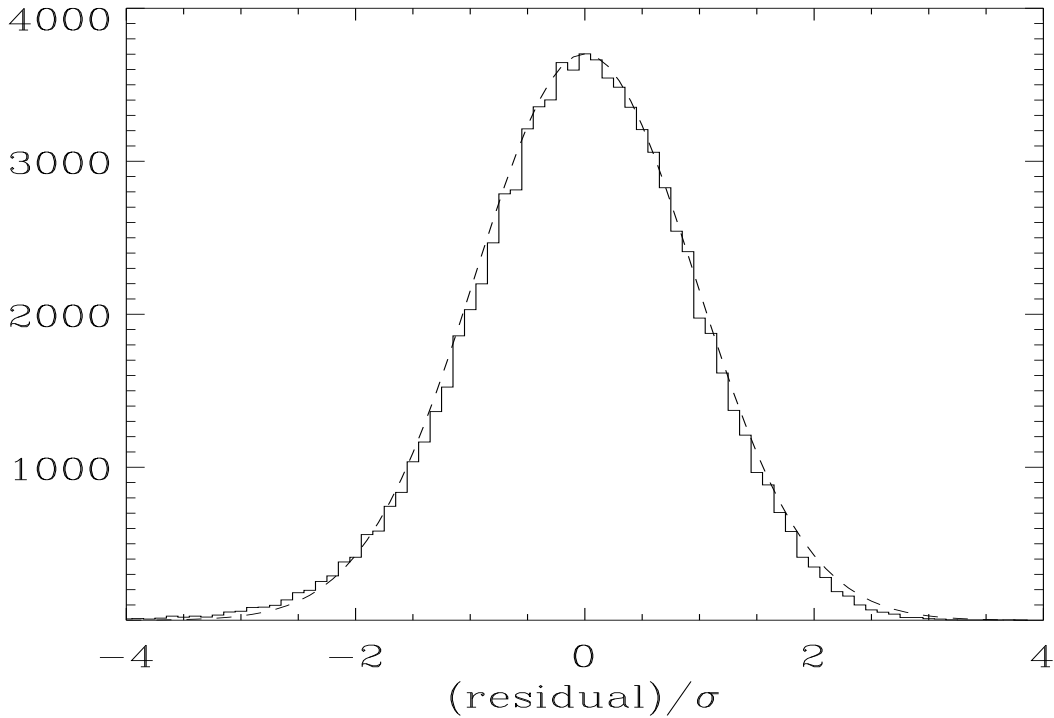}
\end{center}
\caption{A histogram of the
 residuals of sky lines relative to the model sky, in
units of the expected RMS.  Points contributing to this distribution are
drawn only from regions with intensity more than twice the background
between OH lines.  The smooth curve is a gaussian of width 1, showing
that sky subtraction is working to the Poisson level. } 
\end{figure}

\begin{figure}[t]
\begin{center}
\includegraphics[width=6in,height=5in]{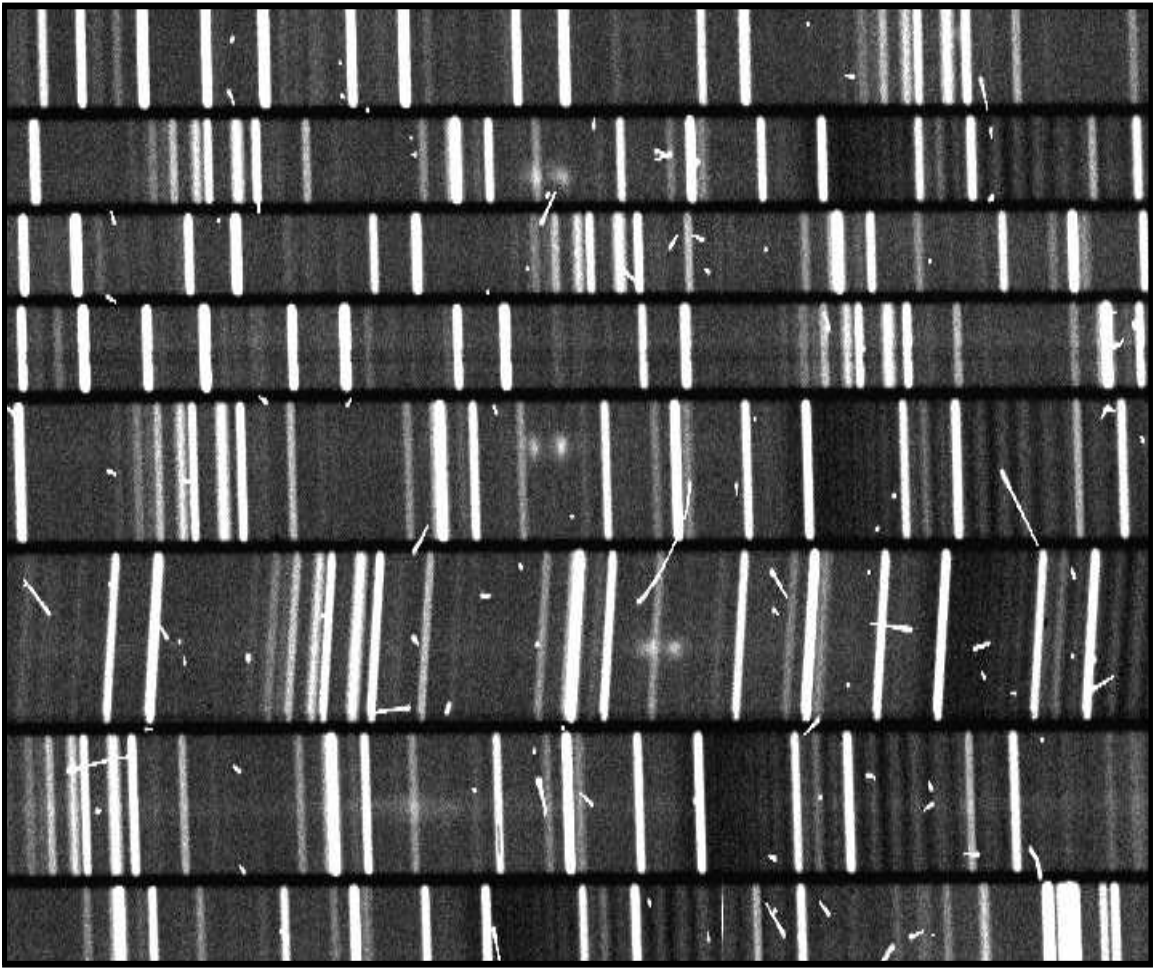}
\end{center}
\caption{A 700 by 300 pixel section of the coadded raw frames of one
mask, with total integration time of one hour.  The tilted sky lines
indicate the tilt of the individual slitlets.  Cosmic rays are
prominent, but note as well
the sets of double emission lines in the spectra. } 
\end{figure}

\begin{figure}[t]
\begin{center}
\includegraphics[width=6in,height=5in]{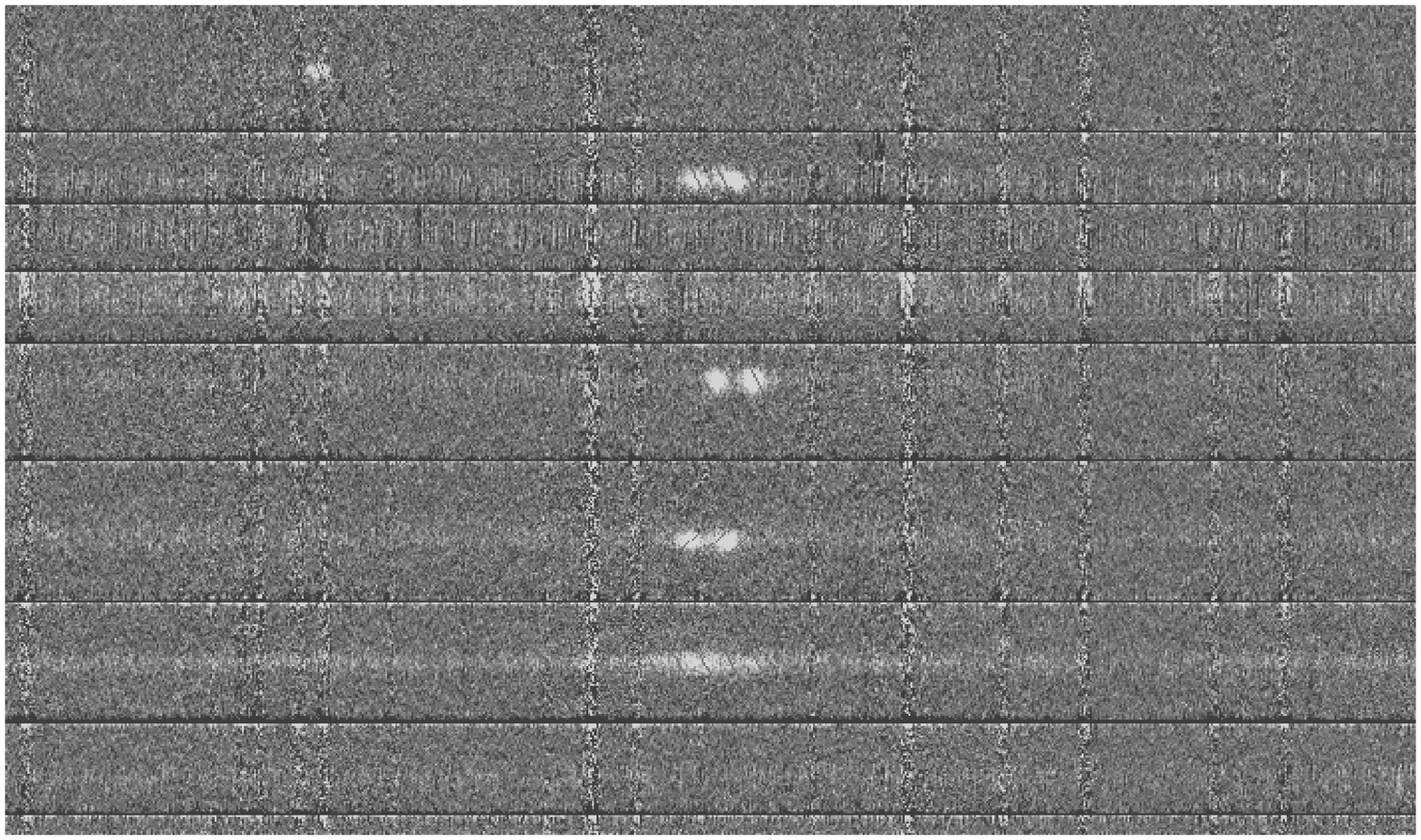}
\end{center}
\caption{A same small piece of a single mask shown in figure 4, 
now having passed through the
reduction pipeline.  The spectra are now aligned in wavelength, and the
noisy vertical lines are residuals of the OH forest after sky
subtraction. The plot does not capture the fact that these lines have
mean zero and noise consistent with their Poisson fluctuation.  The
double emission lines on the four objects in the middle of the plot
are [OII] 3727 $\AA$ 
lines at $z\sim 1$ (the split of the doublet is 220 km/s).  
  This is a modest group of galaxies, one
of the thousand or so expected within the DEEP2 Redshift Survey.} 
\end{figure}

\section{The 1-Hour and 3-Hour Surveys}  

The DEEP2 project is actually subdivided into two surveys, 
the 1HS (one hour [integration time]
survey) and the 3HS (three hour survey).  The 1HS covers a larger area
to a shallower depth and  will require 90 nights on Keck for
execution, while the 3HS covers a small area to greater depth and will
require 30 nights of Keck time.  The two 
subsurveys will serve somewhat different purposes, and we
describe them in more detail here.

{\bf The 1-Hour Survey: 55,000 galaxies to $R_{AB}$ = 24.1 over 3.5
square degrees}.  The 1-Hour Survey (1HS) is the backbone survey that
will provide linewidths, spatial distributions, clustering, and
environmental statistics for tens of thousands of brighter galaxies.
Its limiting magnitude is $R_{AB}$ = 24.1 mag, which corresponds
to $M^* + 0.5$ at $z = 1$ for an LCDM universe.  Experience to date shows
that to achieve the goal of 55,000 measured redshifts will require
targeting a total of 65,000 galaxies.  Because it will cover a
much larger volume than the 3HS ($\sim 6 \times 10^6h^{-3}$ Mpc vs. a
few $ \times 10^5 h^{-3}$ Mpc), the 1HS will be greatly superior for
studying rare objects such as AGN and clusters of galaxies.
%Conveniently, the space observatories best suited for observing those
%objects -- XMM-Newton and Chandra -- have fields of view comparable to the
%short dimension of a typical 1HS field (30 \arcmin).  Most DEEP2
%objects observed by SIRTF or GALEX in the Groth strip will come from the 1HS,
%and if HST ACS imaging over wide fields is obtained, the 1HS will
%provide the bulk of the complementary spectroscopy.
Spectra and CFHT photometry will be available for every object in the
1HS, and linewidths will be measured for roughly half of those objects
with redshifts.  
However, the 1HS
exposure times are short enough that we will tend to miss redshifts
and linewidths for faint red {\it absorption-line} spectra, for which we
would need more photons to get a redshift than bright, emission line 
galaxies.  Covering such gaps in the 1HS is one goal of the 3HS (see below).

{\bf The 3-Hour Survey: 5,000 galaxies to $I_{AB} = 24.5$}.  The 3HS
is a survey-within-a-survey that complements and completes the 1HS in
several ways.  The goals of the 3HS are to: (1) obtain redshifts and
linewidths for absorption-line spectra to the same limiting redshift
as the 1HS---crucial for measuring the number density of {\it large}
galactic halos, since the deepest galactic potential wells are
expected to harbor absorption-line galaxies like the local giant
ellipticals; (2) aided by HST imaging, measure complete
structural parameters and morphologies of galaxies at $z \sim 1$ down
to $\sim M^* + 1.5$; and (3) reobserve a subset of 1HS objects with
higher S/N to obtain definitive completeness statistics for the 1HS.

\section{SCIENCE GOALS OF THE DEEP2 Redshift Survey }

Our team's principal science goals for the DEEP2 Redshift Survey
center on galaxy formation, galaxy biasing, and fundamental cosmology,
as detailed below.  However, a broad array of projects will be made
possible by the DEEP2 data archive, including studies of AGN evolution
in the early universe; exploration of the origin of the cosmic
infrared background; development of total-matter maps to compare to
weak lensing surveys; calibration of photometric redshifts;
measurements of the cosmic evolution of star formation and chemical
abundances; and measurements of the number density of very distant
galaxies at $z>5$. 
% Many of these projects will require complementary
%observations that may only be obtained from space to be effective.
The principal science goals of the ground-based portions of DEEP2
Redshift Survey are: 

%\medskip
\noindent{\bf Goal 1: Determine the characteristics of
galaxies at $z \sim 1$ and their dependence on environment; e.g.
measure the evolution of the ``structure function" of galaxies with
redshift.}  DEEP2 will measure a wide variety of parameters of the
observed galaxies: not just colors, magnitudes, and redshifts, but
also in many cases linewidths or rotation velocities, equivalent
widths of emission lines (and thereby such parameters as metallicity
and O[II]-derived star formation rates), the ages of stellar
populations, etc., etc.  The distributions of and correlations between
these parameters, along with their evolution to the present epoch,
will provide strong constraints on models of galaxy formation and
evolution, whether semi-analytic (e.g. Refs.~\citenum{benson02,somer02}) or based on N-body simulations
(e.g. Refs.~\citenum{mathis02}, \citenum{murali02}).  An example of this sort of DEEP2 science which is critically
dependent on combining ground-based spectroscopy with space-based
imaging is the measurement of the ``structure function'' of galaxies.

In general a set of two parameters, such as a scale radius and
internal velocity, are sufficient to characterize the structure of a
typical dark matter halo \cite{bullock01,nfw97}.
Current theories for galaxy formation predict different evolving
relationships between these dark halo parameters and observable galaxy
properties such as luminosities (e.g. Refs.~\citenum{mmw98,somer02}.
%; these theories can be tested by insisting that the two
%sets of structure data match.
%
The ``structure function'' is a compact way of characterizing these
relationships at any epoch.  Each galaxy can be represented by its
radius, luminosity, and internal velocity---a point in so-called
``Fundamental Plane'' space.  The co-moving number density of galaxies
in this space is what we term the ``structure function,'' by analogy
with the familiar luminosity function.  A 1-d projection of the
structure function onto the luminosity axis generates the luminosity
function, while 2-d projections onto planes generate the Tully-Fisher
and $D_n$-$\sigma$ functions.  A major goal of DEEP2 is to measure the
full structure function at $z \sim 1$ in a wide variety of
environments (from near-voids to clusters) and compare it to today's.
%The relationship between changes in the structure function of visible
%galaxies and the evolution of dark halos provides a powerful test of
%galaxy formation theories.
DEEP2 will measure two parts of this function well: internal
velocities of galaxies can be measured spectroscopically down to $\sim
25$ \kms (limited by the resolution of our spectra); while the
ground-based $BRI$ imaging can be used to measure the (rest-frame $B$)
luminosities of the target objects.  However, from the ground we can
derive radii only for the largest, most spatially extended (or
relatively nearby) galaxies.  

Thus, measurement of the full structure function will only be possible
with the addition of high-resolution space-based images (adaptive
optics from the ground is still limited to comparatively tiny fields
of view).  The ACS instrument recently installed on HST is
ideal for this task.  
%In fact, the DEEP2 team participated in a HST
%Treasury proposal in the last cycle to obtain such observations (along
%with a wide variety of others); in the wake of its failure to be
%approved, we plan to submit a smaller proposal in the next cycle only
%to obtain ACS imaging in the DEEP2 Groth field.
%However, measuring the structure function goes well beyond simply
%analyzing the ACS images; we must combine the ACS and DEEP2 results,
%characterizing in full the completeness and systematics of each, to
%have a useful measurement.  Extensive simulations will be required,
%etc.; this goes well beyond what HST funding would cover.This project
%will {\bf only} be possible if we can integrate the results of ACS
%imaging with the data obtained by DEEP2; no other past or planned
%survey can do anything like it.
With measurements of radii from HST, the 3HS would provide complete
structural data for about 5,000 galaxies, sufficient to allow
subdivision by galaxy type and environment; the 1HS objects in the
same fields would provide $\sim 5,000$ more.  Most galaxies of the 1HS
will lack accurate radii (unless HST imaging were eventually obtained
for all the DEEP2 fields) but will still provide excellent data on the
luminosity function, $N(L)$, the velocity function, $N(v)$, and the
Tully-Fisher function, $N(L,v)$.  If the cosmic input parameters
($\Omega_m$, $\Omega_{\Lambda}$, etc.) are known from other
measurements, these functions $N(L)$, $N(v)$, and $N(L,v)$ will also
provide useful tests of the baryonic infall physics of galaxies.  

In addition, Palomar K-band imaging will be performed over the full 1HS survey
region, which will provide invaluable information  on stellar masses
for  galaxies at $Z \sim 1$.  The K-band photometry, combined with
linewidths and HST determined sizes, will also allow us to determine
stellar to total mass ratios. % particularly in the 3HS survey. 
 This is an
important quantity, but has hardly been attempted at high redshift; DEEP2
should be able to do it for many galaxies.

%Other space-based observations would allow us to change the measure of
%luminosity used in defining the structure function, providing novel
%measurements of galaxy characteristics to be compared to theoretical
%predictions.  For instance, SIRTF/IRAC observations will measure
%rest-frame near-infrared luminosities at $z\sim 1$, providing an
%excellent measure of the total stellar mass in those objects, 
%(with
%less dependence on stellar populations that rest-frame $B$) 
%a quantity easier to model than optical luminosity.  GALEX and
%SIRTF/MIPS 
%star-formation rates would provide instead the luminosity coming
%from new-born stars, related intimately to the mergers and feedback
%that are critical to galaxy formation models.  

%\medskip
\noindent{\bf Goal 2: Measure the two-point and
higher-order correlation functions of galaxies at $z \sim 1$ as a function
of other observables.}  In almost all models of structure formation
(e.g., Ref.~\citenum{white87}), galaxies are born as highly biased tracers
of the mass distribution, but their bias diminishes with time. Spiral
galaxies today appear to be weakly biased, if at all, while the
clustering of $z \sim 3$ Lyman-break galaxies requires a large bias for any
reasonable cosmological model \cite{porciani02,bullockws02}.
%(Porciani \& Giavalisco 2002, Bullock,
%Wechsler \& Somerville 2002).  
Galaxies at $z \sim 1$ should have an intermediate
degree of bias, with readily observable consequences. 
With sufficiently dense sampling, determining the
higher-order clustering properties of galaxies can yield direct measurements of
their biasing \cite{fry96,scocci98}.
%.  For example, in linear biasing models [where $(\delta
%\rho/\rho)_{gal}$ is assumed to be some constant $b \times (\delta
%\rho/\rho)_{mat}$], the mean ratio of the skewness of the density
%distribution function to the square of its variance has an expectation
%value that scales as $1/b$ 
%(Fry~1996; Scocciamarro et al.~1996).  
It will be possible to subdivide the 1HS sample as a function of galaxy type, luminosity, etc.
and measure the biasing for each sample both in an absolute sense and
compared to the other samples.  This and other, more
sophisticated measures  will be explored as part of DEEP2.  
%This
%work is indirectly complementary to space-based studies; with GALEX or
%SIRTF observations in DEEP2 fields, for instance, we can measure
%biasing as a function of stellar mass or star formation rate rather
%than linewidth or luminosity.
The 1HS survey is designed to provide a fair sample volume for analysis
of LSS statistical behavior, particularly for clustering studies
on scales $ < 10h^{-1} $ Mpc. 
 The comoving volume surveyed in the 1HS program will
exceed that of the LCRS survey \cite{shectman}, a survey 
 which has proven to be an 
outstanding resource for low redshift studies of LSS. 
%When the data is in hand, we plan a number of scientific analyses, with
%major programs listed below.
%Each of the four DEEP surveys will enclose a densely sampled 
%comoving volume of approximately
%$500 \times 60 \times 8 h^{-3}$ Mpc$^3$ (in an Einstein--de Sitter cosmology).
%The densely sampled portions of the survey will thus be approximately two-dimensional,
%but the shortest dimension does exceed the correlation length of the galaxy
%clustering.  The outrigger fields will yield information in 
%a dilutely sampled cone that is well suited for detection of large scale
%filaments.  We have a number of science goals in mind for this database:

\noindent{\bf Goal 3: Determine the evolution of the abundance
of dark matter halos and clusters 
as a function of internal velocity, $N(v,z)$}.  
By measuring the linewidths of parent dark matter halos
from the galaxies visible within them (as per Goal 1 above), we can
use the dark-halo abundance as a function of internal velocity and
redshift, $N(v,z)$, to perform a classic cosmological test.  It is
well known that the volume element $dV/dzd\omega$ (where $\omega$ is
solid angle) strongly depends on the input cosmological parameters,
notably $\Omega_m$ and $\Omega_{\Lambda}$.  Thus, the apparent number
of objects with a given linewidth versus redshift is a sensitive test
of the volume element---provided the co-moving number density of
those objects is known.  In practice, the poorly-known evolution of the
number density $N(L)$ of galaxies has stymied this test.  However, if
we have measured real potential-well depths, we can bypass galaxies
and count the more easily simulated dark halos directly.
% characterized by their
%linewidth $v$.  
This work will require us to study a significant
fraction of our galaxies with the high resolution of HST to ensure
that these objects are morphologically simple, and thus that their
linewidths provide real information about the potential wells of
galaxies.
Newman \& Davis \cite{nd00,nd02} showed that the degree of evolution in the
comoving number density of
galaxy sized halos at fixed velocity 
%evolves modestly between z=0 and
%z=1; the amount of that evolution turns out to be 
is almost totally independent of cosmology.  The observed abundance of
such objects, $dN(v)/dz$, thus measures the volume element of the
expanding Universe
%, which varies by a factor of 3 amongst standard cosmological
%models 
and gives us a powerful handle on the cosmic geometry. 

In contrast, the time evolution of the comoving number density of
groups and clusters of galaxies is exponentially sensitive to certain
combinations of cosmological parameters -- much stronger than the
differences in volume amongst models.  Thus, the detected abundance of
clusters provides a separate, very powerful probe of the cosmological
parameters (e.g. Ref.~\citenum{haimanmh}).  
%DEEP2 will include hundreds of groups and dozens of
%rich clusters within the survey volume.  
Based on tests with mock catalogs from simulations, we expect to
measure the abundance of clusters as a function of their velocity
dispersion -- which can be predicted directly from models of the dark
matter distribution -- down to a dispersion of $\sim 400$ \kms at $z
\sim 1$ \cite{newman02,marinoni02}.  This provides
another test of the cosmology, complementary to the one provided by
counts of galaxies.

Both techniques can place constraints not only on spatially curved
models containing matter and possibly a simple cosmological constant,
but also on flat ``quintessence'' models in which the dark energy is
assumed to consist of an active field with an effective equation of
state $P=w\rho$, with $-1 < w < -1/3$ ($w=-1$ for a cosmological
constant; cf. Ref.~\citenum{wang}).  The equation of state
parameter $w$  can only be measured via global
cosmological tests; prospects for constraining it from CMB analyses
alone are relatively poor \cite{hu99}.

\noindent{\bf Goal 4: Measure redshift-space distortions due to 
peculiar velocities at  $z \sim 1$.} The clustering of galaxies is
inherently isotropic in space, with no preferred orientation toward or
away from the Milky Way.  The observed redshift-space clustering of
galaxies, however, is distorted by peculiar velocities, producing features
such as the so-called ``fingers of God'' on virialized scales and a flattening
of structure on larger scales.  DEIMOS will deliver highly precise redshifts,
allowing both of these effects to be readily
detectable in our maps (see Ref.~\citenum{coil01} for details).   

A variety of statistical tools have been developed to extract pair-
and object-weighted velocity dispersions as well as measures of the
flattening on larger, non-virialized scales.  Analysis of these
quantities will give us strong measures of a degenerate combination of
$\Omega_m$ at $z\sim 1$ and the bias of the galaxy distribution.  This
degeneracy can be broken by comparison to the same statistics at $z
\sim 0$.  By these analyses we can get a strong handle on the bias of the
galaxy distribution and can study the success and failures of our
paradigm of structure formation.  With DEEP2 we can construct at
$z\sim 1$ the detailed statistics of velocity fields that have only
recently been possible at $z\sim 0$ \cite{peacock01}~!.

\section{DATA FLOW FOR THE DEEP2 Redshift Survey}

The data rate from DEIMOS will be in excess of 1 Gbyte/hour, 
necessitating automated  reduction and analysis tools.  %A pipeline code
%based on the SDSS spectral code of Schlegel and Burles 
%has been adapted to our needs and now operational.  
We have therefore developed a spectral reduction pipeline for DEIMOS
based upon the remarkable pipeline developed by D. Schlegel and
S. Burles for the SDSS spectroscopic survey.  We are extremely
indebted to them for allowing us to study their code and extract its
core elements.  As shown above, we are routinely achieving Poisson
limited sky subtraction even amidst the OH forest, which is made
possible by careful attention to the 2-d wavelength solution within
each slitlet and by the use of b-splines for a precision fit of the
sky spectrum within each slitlet.  Details of this spectral reduction
pipeline will be provided elsewhere.

The code for reduction to 
1d spectra is a completed package at this writing, though still being
revised.  Automated
redshift determination and rotation curve analysis are in some ways
easier tasks; we have begun 
adapting D. Schlegel's 1d spectroscopic analysis code to our puroises.
The spectra at all stages of reduction are stored as
FITS files, while the photometric and 
spectroscopic databases for the project are stored as FITS binary
tables, which become IDL structures when read from disk.
Thus the pipeline code and database management are fully integrated
and quite easy to use.  

Reduction of a full mask of calibration files
and 3 science frames requires $\sim 6$ hours on a 1.5GHz linux
workstation; the UCB group has acquired a modest Beowulf cluster which
will be used for all the DEEP2 reductions, and we anticipate no
difficulty in keeping up with the flood of data that DEIMOS will
produce. Although the total raw data will exceed
one Tbyte of compressed FITS files, 
the reduced data will be modest in size by today's standards,
$< 100$ Gbytes.

\section{SCIENCE RESULTS FROM EARLY DEIMOS DATA}

The weather for DEIMOS commissioning in June, 2002, was excellent, but 
the July and August science runs yielded only one useful night out of
the 9 nights assigned.  Fortunately our luck changed for the September
run, during which 29 masks were completed in a 4 night period.  As of 
this writing, we have
collected spectra of more than 4000 high redshift galaxies.
%Only one  of the four July science nights on Keck assigned to the DEEP2
%was useable, and it was plagued by poor seeing and cirrus.  
%Excellent
%data was obtained, however, in early commissioning nights in June,
%during which 
%two masks for the 1HS survey were observed. 
Figure 5 is a demonstration that emission lines are prominent in a substantial
fraction of our target galaxies.  These spectra have sufficient SNR
to estimate the line ratio of the doublet as well as the internal
broadening.  In many of our target galaxies,  we can trace the
rotation curve.  
  The closely associated galaxies of figure 4 form a modest group of
galaxies, and we expect to observe hundreds of such groups over the
course of the survey.  Details on these results, and many others, will
be forthcoming after the close of the 2002 observing season in
October.

\section{CONCLUSIONS}
The flood of data now coming from the SDSS and 2DF  
projects detailing
the local Universe is beginning to be 
complemented by data from the VLT/VIRMOS project and by the DEEP2 Redshift 
Survey 
providing detailed information on the Universe at $z \approx 1$,
thus continuing the revolution in precision cosmology and large-scale 
structure.  These results will rapidly expand in the coming months and
will keep us all extremely busy. 
We intend to share our results with the public and to put our
spectra online in a timely manner.  Further details on the survey can be
found at the web site
http://deep.berkeley.edu/. 

% Begin acknowledgements
\begin{acknowledgements}

{This work was supported in part by NSF grants AST00-71048
and KDI-9872979.   The DEIMOS spectrograph was funded by a grant from CARA
(Keck Observatory), by an NSF Facilities and Infrastructure grant (AST92-2540), by the Center for Particle Astrophysics, and by gifts from Sun Microsystems
and gifts from Sun Microsystems and the Quantum Corporation.
DPF is supported by a Hubble Fellowship.}
\end{acknowledgements}

%INDEX%%%%%%%%%%%%%%%%%%%%%%%%%%%%%%%%%%%%%%%%%%%%%%%%%%%%%%%%%%%%%%%
%\clearpage
%\addcontentsline{toc}{section}{Index}
%\flushbottom

%%%%%%%%%%%%%%%%%%%%%%%%%%%%%%%%%%%%%%%%%%%%%%%%%%%%%%%%%%%%%%%%%%%%%

\end{document}